\documentclass[twocolumn]{aastex63}

\usepackage{color}
\usepackage{soul}
\usepackage{tabulary}
\usepackage{textcomp, gensymb, amsmath}
\usepackage[T1]{fontenc}
\usepackage{float}
\usepackage{appendix}
\usepackage{multirow}
\newcolumntype{K}[1]{>{\centering\arraybackslash}p{#1}}

\shorttitle{Kepler-289 TTVs}
\shortauthors{Greklek-McKeon et al.}

\graphicspath{{./}{figures/}}

\begin{document}

\title{Constraining the Densities of the Three Kepler-289 Planets with Transit Timing Variations}

\correspondingauthor{Michael Greklek-McKeon}
\email{mgreklek@caltech.edu}

\author[0000-0002-0371-1647]{Michael Greklek-McKeon}
\author[0000-0002-5375-4725]{Heather A. Knutson}
\author[0000-0003-2527-1475]{Shreyas Vissapragada}
\affiliation{Division of Geological and Planetary Sciences, California Institute of Technology, Pasadena, CA, 91125, USA}
\author[0000-0002-6227-7510]{Daniel Jontof-Hutter}
\affiliation{Department of Physics, University of the Pacific, 3601 Pacific Avenue, Stockton, CA 95211, USA}
\author[0000-0003-1728-8269]{Yayaati Chachan}
\affiliation{Division of Geological and Planetary Sciences, California Institute of Technology, Pasadena, CA, 91125, USA}
\author[0000-0002-5113-8558]{Daniel Thorngren}
\affiliation{Department of Physics, University of Montr\'eal, 1375 Avenue Th\'er\`ese-Lavoie-Roux, Montr\'eal, QC, H2V 0B3, Canada}
\author[0000-0002-1871-6264]{Gautam Vasisht}
\affiliation{Jet Propulsion Laboratory, California Institute of Technology, 4800 Oak Grove Dr, Pasadena, CA 91109, USA}

\begin{abstract}

Kepler-289 is a three-planet system containing two sub-Neptunes and one cool giant planet orbiting a young, Sun-like star. All three planets exhibit transit timing variations (TTVs), with both adjacent planet pairs having orbital periods close to the 2:1 orbital resonance. We observe two transits of Kepler-289c with the Wide-field InfraRed Camera (WIRC) on the 200'' Hale Telescope at Palomar Observatory, using diffuser-assisted photometry to achieve space-like photometric precision from the ground. These new transit observations extend the original four-year Kepler TTV baseline by an additional 7.5 years. We re-reduce the archival \textit{Kepler} data with an improved stellar activity correction and carry out a joint fit with the Palomar data to constrain the transit shapes and derive updated transit times. We then model the TTVs to determine the masses of the three planets and constrain their densities and bulk compositions. Our new analysis improves on previous mass and density constraints by a factor of two or more for all three planets, with the innermost planet showing the largest improvement. Our updated atmospheric mass fractions for the inner two planets indicate that they likely have hydrogen-rich envelopes, consistent with their location on the upper side of the radius valley.  We also constrain the heavy element composition of the outer saturn-mass planet, Kepler-289c, for the first time, finding that it contains 30.5 $\pm$ 6.9 $M_{\oplus}$ of metals. We use dust evolution models to show that Kepler-289c must have formed beyond 1~au, and likely beyond 3~au, and then migrated inward.

\end{abstract}

\section{Introduction}

The \textit{Kepler} space telescope \citep{Kepler} discovered thousands of new transiting exoplanet systems, including hundreds of multi-planet systems. \textit{Kepler} was decommissioned in 2018, but the science value of the data it provided has yet to be exhausted. Part of this legacy science includes follow-up observations of the most interesting dynamically interacting multi-planet systems. \emph{Kepler} transit photometry provides us with information about the planetary radius, orbital period, inclination, and semi-major axis of transiting planets.  For the subset of dynamically interacting systems, observations of transit timing variations \citep[TTVs;][]{Agol_2005, Murray_2005} also provide complementary constraints on planetary masses, eccentricities, and bulk densities.  We can use this information to characterize the bulk compositions and atmospheric mass fractions of these planets, as well as their dynamical architectures, which in turn provide key constraints on planet formation and evolution models.

For an isolated transiting planet on a stable circular orbit, transits occur at uniform intervals. In multi-planet systems, however, gravitational interactions between planets can cause the transit mid-times to deviate from a linear ephemeris by amounts ranging from minutes to hours. The amplitude of these TTVs depends on the masses of the planets and their orbital elements \citep{Agol_2018}. Planets near first-order mean-motion resonances (MMRs) typically have the largest TTV amplitudes \citep[e.g.][]{Deck_2016} and are therefore the most favorable for dynamical mass measurements.

\textit{Kepler} discovered hundreds of multi-planet systems exhibiting detectable TTVs \citep{Holczer,Hadden_2017}. For many of these systems, the data from the four-year \textit{Kepler} prime mission was sufficient to obtain a large sample of transits for each planet spanning the TTV ``super-period", the timescale on which the TTVs oscillate \citep{Lithwick_2012}. However, the \emph{Kepler} data are unable to provide strong constraints on the dynamical states of some planets with orbital periods longer than $\sim$100 days, where relatively few transits were observed and/or the baseline is shorter than the predicted super-period. In cases where the \emph{Kepler} baseline is longer than the TTV super-period and we can obtain good dynamical mass constraints, TTVs provide us with a unique opportunity to measure the bulk densities of long-period planets, without requiring an extensive multi-year radial velocity (RV) follow-up campaign \citep[e.g.][]{Chachan_2022}. There are currently only 18 transiting planets with a fractional mass uncertainty less than 1/3 (e.g., $>3\sigma$ mass measurement) that have orbital periods greater than 100 days listed in the Exoplanet Archive \citep{exoarchive}. Of these 18, 11 were characterized using TTVs. 

In this study we focus on Kepler-289, a young ($\sim$1 Gyr) Sun-like ($T_{\mathrm{eff}}=5990\pm38$ K) star that hosts three planets with orbital periods of 34.5, 66.1, and 125.9 days, corresponding to period ratios of 1.9:1 for each adjacent planet pair \citep{schmitt}. These period ratios are reminiscent of those of the Galilean moons of Jupiter, which are in a Laplace resonance. We calculate a normalized distance to resonance $\Delta$ of 0.04 for both planet pairs \citep{Lithwick_2012}, indicating that they are likely to exhibit TTVs. This is consistent with the population of near-resonant Kepler systems identified in previous studies, which typically have $\Delta < 0.05$ \citep{Holczer,Jontof-Hutter_2016, Jontof-Hutter_2021}. Unlike most of these sytems, the Kepler-289 planets have period ratios interior to resonance, rather than the more common exterior configuration \mbox{\citep{Fabrycky_2014}.}

Although sub-Neptune planets frequently have outer Jovian companions \citep{bryan_2019,zhu_wu_2018}, these companions are typically located at much larger orbital separations ($\gtrsim1$~au). The relatively close spacing between the outer gas giant and the inner sub-Neptunes in the Kepler-289 system, along with their near-resonant orbital configuration, suggests that this system may have started with a much wider orbital spacing before undergoing migration \citep{Charalambous_2022}. The planets are not currently in resonance, however, and by obtaining improved mass measurements and bulk composition constraints for all three planets we can further examine the unique history of this unusual system.

All three planets in the Kepler-289 system exhibit detectable TTVs in the \emph{Kepler} data, with amplitudes ranging from $\sim0.5-5$ hours \citep{schmitt}. Previous TTV studies of this system were complicated by the fact that the light curve of Kepler-289 is significantly variable ($\sim$3\% amplitude in the \emph{Kepler} bandpass) due to rotational modulation from starspots. This makes it difficult to obtain reliable transit mid-time measurements for the inner and middle planets, which both have relatively shallow transit depths of $\sim$400 ppm \citep{schmitt}. In addition, the outermost planet, Kepler-289c, only transited ten times during the \emph{Kepler} observations, which cover only slightly more than one full TTV super-period \citep{Jontof-Hutter_2021}. This makes this planet an ideal target for ground-based follow-up observations, which allow us to extend the TTV baseline by many years and to obtain improved dynamical mass constraints for all three planets. 

In this study we re-analyze existing \emph{Kepler} photometry of this system and combine it with two new observations of Kepler-289c obtained using the Wide Field InfraRed Camera (WIRC) on the 200" Hale Telescope at Palomar Observatory.  We achieve space-quality infrared photometry of this system by using a beam-shaping diffuser whose properties are described in more detail in \cite{Stefansson} and \cite{Vissapragada}. By observing in the infrared, we mitigate the effects of stellar activity on the transit light curve. In Section~\ref{sec:obs}, we describe each observational data set. In Section~\ref{sec:transit_fits}, we describe our stellar activity correction for the \emph{Kepler} photometry, our reduction of the Palomar transit light curves, and the TTV analysis. In Section~\ref{sec:results} we discuss the results of our analysis, and in Section~\ref{sec:conclusions} we conclude and discuss the implications for these new set of constraints on the Kepler-289 system. 

\begin{figure*}
  \includegraphics[width=\textwidth]{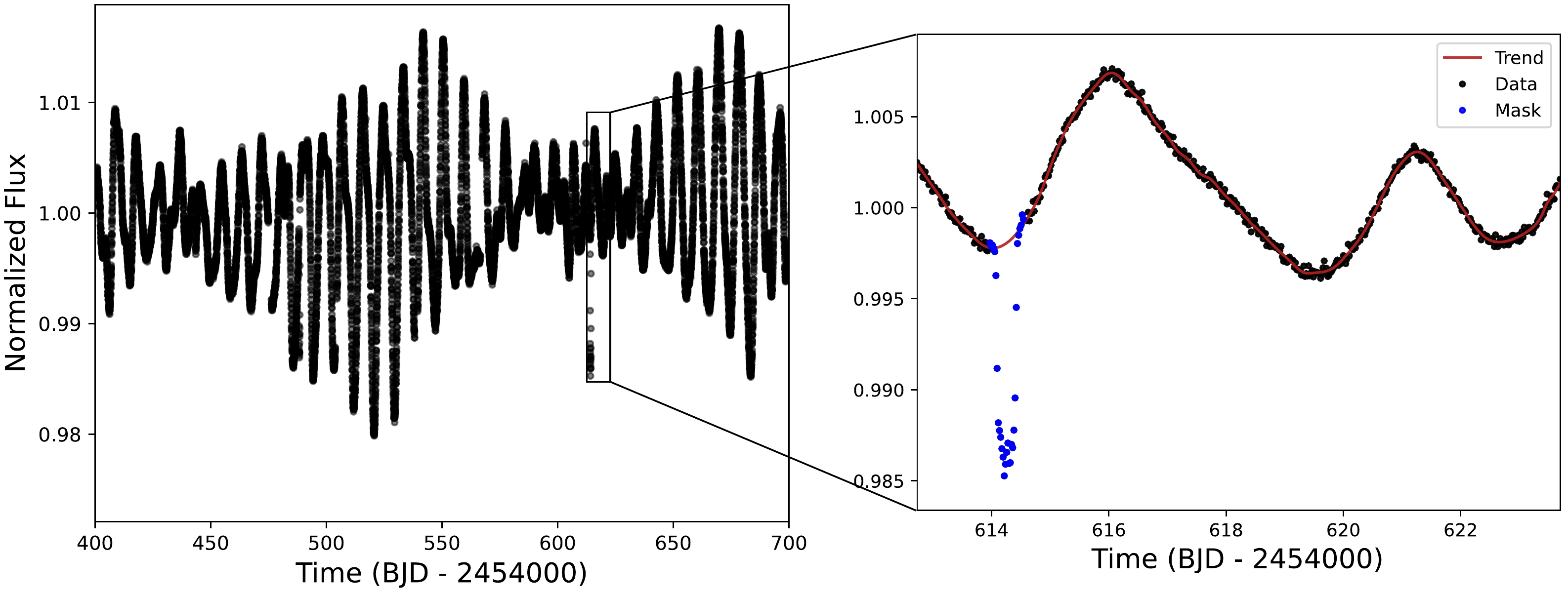}
  \caption{Representative photometry of Kepler-289 from \textit{Kepler} quarter 5, exhibiting significant rotational modulation. A deep transit of planet c, the outermost planet, is easily visible around 614 (BJD-2454000). The right panel shows a zoomed-in view of this transit with the masked region around the transit indicated in blue and the baseline trend we use to remove the stellar variability overplotted in red.}
  \label{Fig 1}
\end{figure*}

\section{Observations}\label{sec:obs}

\subsection{Space-Based Photometry}\label{subsec:space_obs}

The Kepler-289 system was observed continuously during \textit{Kepler} Quarters 1–16, with only long-cadence (30 minute integrations) data available for the first eleven quarters, and short-cadence (1 minute integrations) data in the remaining five. We obtained the Pre-search Data Conditioning Simple Aperture Photometry (PDCSAP) flux from the Kepler-289 postage stamp in Quarters 1–16 from the Mikulski Archive for Space Telescopes (MAST), and we perform our own reduction of the Kepler photometry, described further in the following section. The PDCSAP photometry already has long-term instrumental trends removed from the data using co-trending basis vectors.

There are also \emph{TESS} \citep{Ricker2014} data available in sectors 14, 15, and 41 for Kepler-289, which we downloaded using the lightkurve package \citep{lightkurve} and checked for transits of the outer planet, Kepler-289c. We find that \emph{TESS} observed the same two transits that we observed with WIRC, but the \emph{TESS} photometric precision is significantly worse. We therefore exclude the \emph{TESS} data from our subsequent analysis. 

\subsection{Ground-based WIRC Photometry}\label{subsec:wirc_obs}

Kepler-289c's 126-day orbital period makes it a challenging target to observe from the ground, as the transit has a total duration of $\sim$8 hours and there are typically no opportunities to observe full transits of this planet at Palomar. We obtained two partial transits of Kepler-289c in \textit{J} band using the Wide-field InfraRed Camera (WIRC) instrument at the prime focus of the 200" Hale Telescope at Palomar Observatory. Our first observation was taken on on UT 24 August 2019 with a baseline spanning slightly more than half of the total transit duration, beginning approximately an hour before the transit center and ending approximately an hour after the end of egress. We observed a second partial transit on UT 17 September 2021, beginning $\sim$2.5 hours pre-ingress and observing until just past the mid-transit time. 

WIRC has an 8.7$\arcmin\times8.7\arcmin$ field of view, ensuring that there are at least 10 comparison stars with magnitudes comparable to that of Kepler-289 visible in the same field. We utilized a custom near-infrared beam-shaping diffuser for these observations, which creates a top-hat PSF with a full width at half maximum of 3$\arcsec$. This diffuser improved our observing efficiency and mitigated time-correlated noise from PSF variations, increasing our observing precision to levels comparable with space-based infrared photometry for a star of this magnitude \citep[$J=12.9$;][]{Stefansson, Vissapragada}. We  minimized the time-correlated noise contribution from flat-fielding errors by utilizing custom guiding  software, which limits the pointing drift over the night to a few pixels \citep{Zhao_2014}. This software guides on science images by fitting 2D Gaussian profiles to comparison stars and determining guiding offsets on each image. We observed on both nights with an exposure time of 25 seconds, which we co-added to a total exposure time of 50 seconds. On both nights we achieved a guiding stability of less than two pixels (0$\farcs5$) throughout the night. On the first night, our observations began at airmass 2.5, reached a minimum airmass of 1.02, and continued until airmass 2.5. On the second night, our observations began near airmass 1.0 and continued until airmass 2.5 when the target set.

\begin{figure*}
  \includegraphics[width=\textwidth]{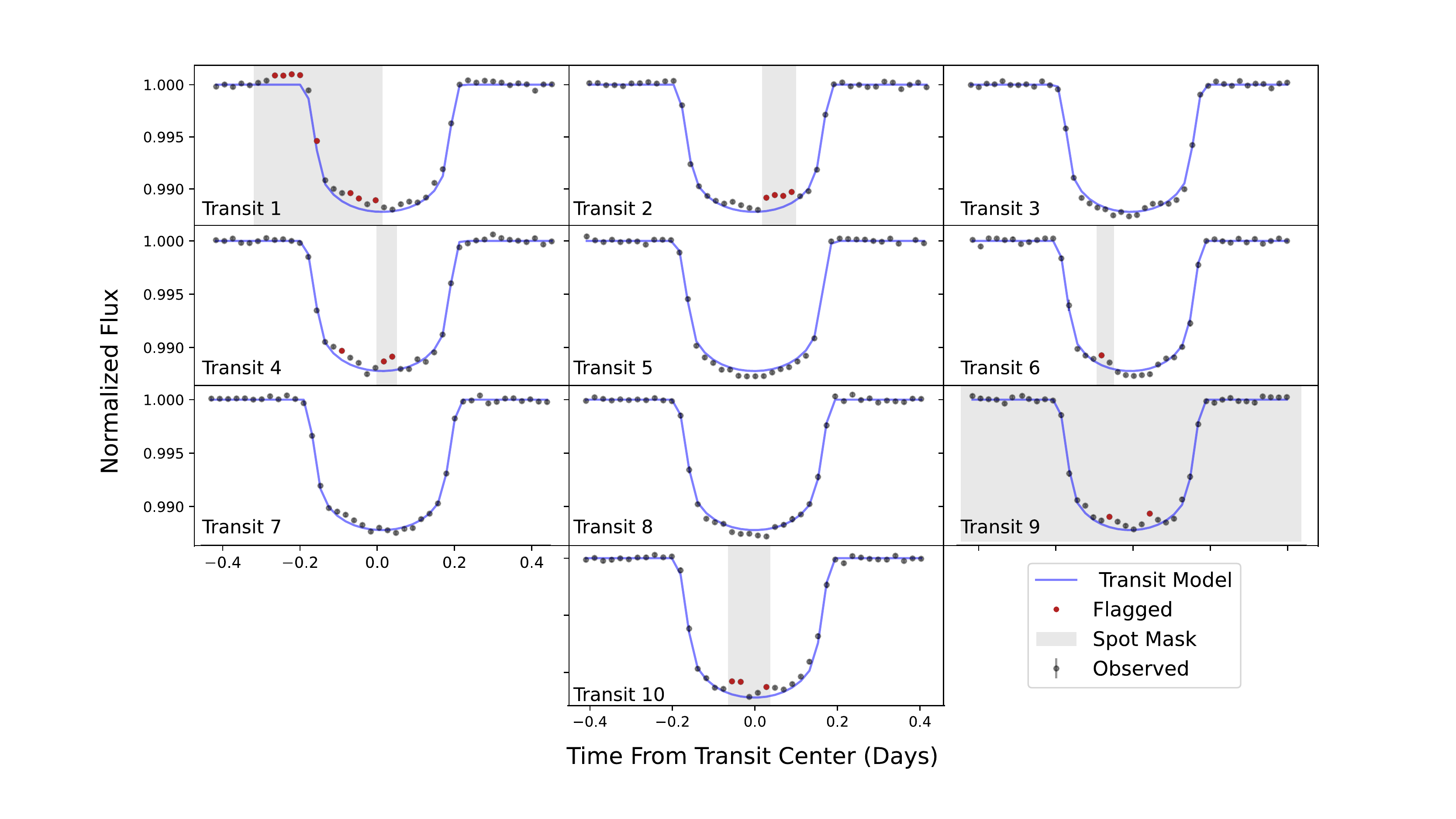}
  \caption{\emph{Kepler} transit photometry of Kepler-289c. Our initial masks from visual inspection of the transits are shown as grey shading, the final best-fit transit model from the stacked transit profile is overplotted as a blue curve, and our final 3$\sigma$ outlier masks are shown as red points. We binned the short-cadence data to match the long cadence data (30 minute bins) for this figure. }
  \label{Fig 2}
\end{figure*}

\section{Data Analysis}\label{sec:transit_fits}

\subsection{Kepler Transit Fits and Stellar Activity Correction}\label{subsec:kepler_transits}
The Kepler simple aperture photometry already has instrumental variations removed, but there is significant astrophysical variability in the light curve of Kepler-289 due to rotational modulation from starspots. This variability is on the scale of $\sim$3\%, while the transits of the inner two planets have depths of approximately 0.4\%, and the outer planet $\sim$1\%. To obtain the highest possible precision on the individual \emph{Kepler} transit times, we first need to remove the astrophysical variability.  

We began by using the transit mid-times reported in \cite{Holczer}, or \cite{Rowe_2014} where mid-times are not available in \cite{Holczer}, to mask all of the data within $\pm$1.5 transit durations (for the relevant planet) of each transit mid-time. We then broke up the light curve into segments, using any gap in the data longer than a day as a break point. For each segment, we fit the long-term trends in flux using a cubic B-spline with knots defined by a smoothing factor of $m + \sqrt{2m}$, with $m$ being the total number of data points \citep{spline}, and interpolated the trend over the masked transit points. We then divided out this trend to flatten the light curve. We show a representative section of the original \textit{Kepler} light curve from quarter 5, displaying significant astrophysical variability, along with a close-up of the trend and mask we used, in \autoref{Fig 1}. 

After removing the stellar variability, our next step was to fit the phased transit profiles for each planet. We used the transit mid-times from \cite{schmitt} to phase up the individual transits for each planet. For the inner and middle planets, which have more shallow transits, no additional steps were required to correct for stellar activity in the light curve for fitting. For the outer planet, which has a  relatively deep transit, we noted that there appear to be ``bump"-like features consistent with starspot crossings \citep[e.g][]{Wolter_2009, davenport2014using} in many of the individual transit light curves. These features were most readily evident in the second, fourth, sixth, and tenth \emph{Kepler} transits, and we masked them in our initial transit fits. The ninth transit was of generally poor quality due to instrumental effects present in the \emph{Kepler} photometry and did not have a well-constrained transit mid-time in the Holczer catalog, so we excluded it from the stacked transit profile. The first half of the first transit also appeared to be affected by uncorrected instrumental effects, so we masked it as well. After defining this initial set of masks, we fit the phased transit profile. We then removed our initial masks for the likely spot crossings and instead flagged all points that were more than 3$\sigma$ away from the best-fit stacked transit model. We found that the majority of the outliers flagged were indeed in the same regions of the light curves as our original spot masks. We use this updated 3$\sigma$ outlier mask in the final joint-fit between the phased \emph{Kepler} data and Palomar data. Masking these spot crossings and instrumental outliers allowed us to obtain an improved constraint on the transit depth and ensured that the transit mid-time measurements were not biased. In \autoref{Fig 2} we plot all of the individual \emph{Kepler} transit observations of Kepler-289c, with our initial and final spot masks over-plotted for reference.

\begin{figure*}
  \includegraphics[width=\textwidth]{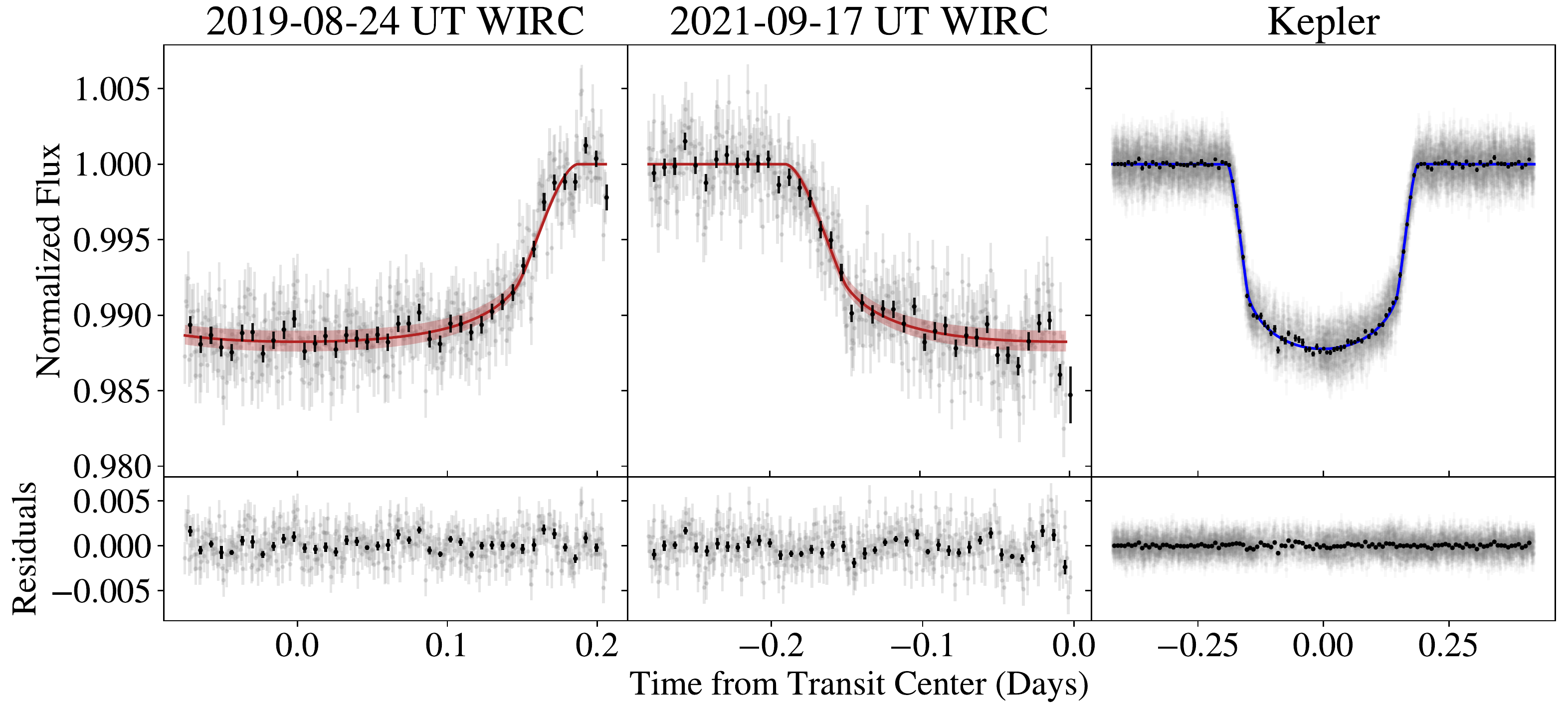}
  \caption{The left panel shows the first partial transit observation in $J$ band from Palomar/WIRC in 2019, middle panel shows the second partial transit observation from WIRC in 2021, and right panel shows the phased \emph{Kepler} transit. The red lines in the left and center panels are the best-fit transit models for the WIRC light curves, and the red shaded region is the 1$\sigma$ confidence interval on that model. The blue line in the right panel shows the best-fit transit model for the \emph{Kepler} data, along with blue shading for the 1$\sigma$ confidence interval. All three of these transit data sets are fit jointly with \texttt{exoplanet}, where the \emph{Kepler} transit profile has separate depth and limb-darkening parameters, and all other parameters are shared aside from transit mid-times.}
  \label{Fig 3}
\end{figure*}

We modeled the transit light curves for the inner two planets using the exoplanet light curve modeling package \texttt{BATMAN} \citep{batman}. We fit for the planet to star radius ratio, orbital inclination, semi-major axis to stellar radius ratio, and quadratic stellar limb-darkening parameters. We used the results reported in \cite{schmitt} as initial guesses for the planet parameters, and sampled their posterior distributions using the affine-invariant ensemble Markov chain Monte Carlo fitting package \texttt{emcee} \citep{emcee}. We run the sampler with 50 walkers for $5.5\times\mathrm{10}^{5}$ iterations and discard the initial $5\times\mathrm{10}^{4}$ steps as burn-in. We checked for convergence by verifying that the number of iterations for each parameter in our MCMC chain exceeded 50 autocorrelation lengths as calculated by \texttt{emcee}.  After fitting the transit shape parameters, we retrieved the ``best-fit" values from a $\chi^2$ minimization on the 50$^{th}$ percentile  parameter set in our posterior chain, and then used the 16$^{th}$ and 84$^{th}$ percentile parameter values from the posterior distribution as our 68\% $\pm$1$\sigma$ confidence interval. We then fit each transit of the inner and middle planets in the \emph{Kepler} photometry with each parameter from the best-fit transit profile fixed except for the transit time, to re-derive transit times for both planets. These times are listed in \autoref{times_table}, and the stacked transit profiles for Kepler-289b and d are shown in \autoref{Kepler289b_transit_plot}. The radii we retrieve for Kepler-289b and Kepler-289d are slightly different than the radii reported in \cite{schmitt} (within 2$\sigma$ for the inner planet, and 1$\sigma$ for the middle planet), which we attribute to our different stellar activity removal process. 

\bigskip

\begin{figure*}
\begin{center}
  \includegraphics[width=\textwidth,height=\textheight,keepaspectratio]{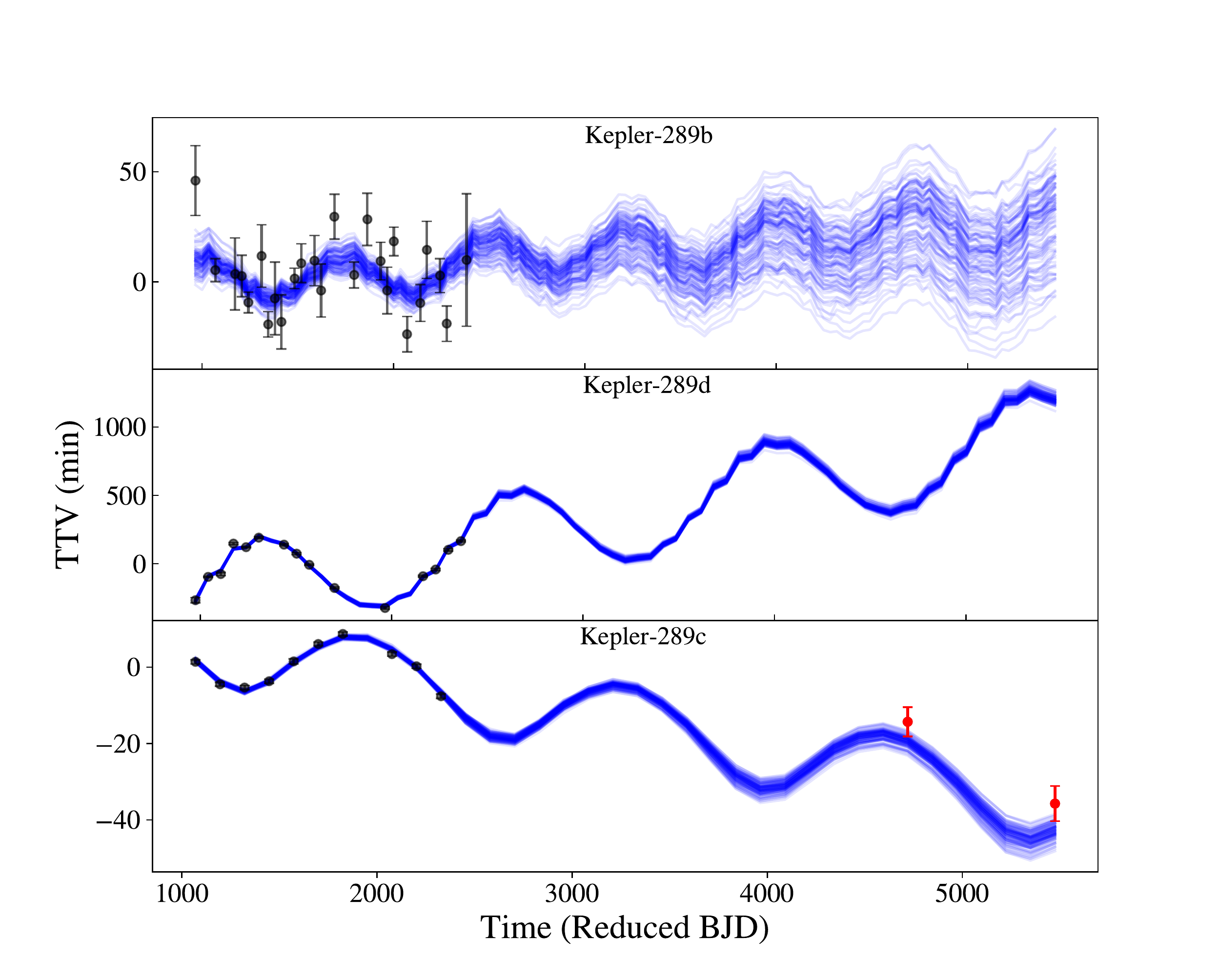}
  \caption{Observed TTVs from \emph{Kepler} (black points) and Palomar/WIRC (red points) for each of the Kepler-289 planets, along with 100 random posterior draws from our TTV model (blue curves). The inner, middle, and outer planets of Kepler-289b, Kepler-289d, and Kepler-289c respectively have mean orbital periods of 34.55, 66.06, and 125.85 days. The planets are labeled in alphabetical order of their discovery in the \emph{Kepler} data. The Palomar transit observations extend the \emph{Kepler} TTV baseline by more than 7.5 years and help refine the TTV super-period originally inferred from the \emph{Kepler} data to $\sim$1344 days. The reduced BJD time is BJD-2454000.}
  \label{ttvplot}
 \end{center}
\end{figure*}

\subsection{Palomar Photometry and Transit Fits for Kepler-289c}

We dark-subtracted, flat-fielded, and corrected our WIRC images for bad pixels and hot pixels using the methods described in \cite{Vissapragada}. We then used aperture photometry with the \texttt{photutils} package \citep{larry_bradley_2020_4044744} to determine the relative fluxes of Kepler-289 and a set of comparison stars in each image.  We tested every aperture size from 5-25 pixels and compared the root-mean-square deviation of the final light curves for each aperture radius. We found that the optimal aperture radius was 9 pixels ($2\farcs25$) for the first night , and 11 pixels ($2\farcs75$) for the second night. We calculated light curves for every star in the image whose cumulative PSF had a signal-to-noise ratio of 100 or more relative to the sky background, and selected the ten stars with the lowest median absolute deviation relative to Kepler-289's flux as our final set of comparison stars. The full set of comparison stars are not identical across our two WIRC observation nights, due to slightly different detector positioning on the sky. We estimated and subtracted the sky background from each star using an uncontaminated sky annulus with inner and outer radii of 25 and 50 pixels, respectively.

We then constructed a systematic noise model that accounts for changes in Kepler-289's flux due to changing airmass, atmospheric seeing, telescope pointing, and other instrumental effects.  This systematics model is a linear function of the ten comparison star light curves, the airmass, the PSF width, the distance moved on the detector relative to the initial centroid position, the local sky background, and the mean-subtracted times.
For more details on the WIRC data reduction pipeline and instrumental noise model, see \cite{Vissapragada}.  The WIRC data reduction and light curve modeling software are publicly available online at \href{https://exowirc.readthedocs.io/en/latest/?}{readthedocs}.

We jointly fit the \emph{Kepler} and Palomar transit light curves using \texttt{exoplanet}. The \texttt{exoplanet} package \citep{exoplanet:joss} uses \texttt{starry} \citep{2019} to quickly calculate accurate limb-darkened light curves and \texttt{PyMC3} \citep{exoplanet:pymc3} to explore the posterior distribution of modeled parameters with No-U-Turn Sampling \citep{hoffman2011nouturn}, an extension of the Hamiltonian Monte Carlo algorithm that utilizes first order gradient information to step through posterior distributions quickly and efficiently. We allowed the mid-time of each transit to vary individually in this fit using the \texttt{TTVOrbit} model, with the stacked \emph{Kepler} transit profile shifted to a reference time one orbital period before the first WIRC transit observation. In our joint \emph{Kepler}/WIRC transit fit, we use global parameters for the planetary orbital period, impact parameter, and semi-major axis to stellar radius ratio.  We fit for separate planet to star radius ratios and quadratic limb darkening parameters in each bandpass. We also fit for a jitter parameter describing the excess noise added in quadrature to the photon noise in each bandpass. In the left and middle panels of \autoref{Fig 3}, we show the final WIRC transit light curves and their residuals.
In the right panel, we show the stacked \textit{Kepler} light curve. 
Our WIRC light curves result in a transit mid-time precision of $\sim$4 minutes, compared to a median mid-time precision of 0.55 minutes from the space-based \textit{Kepler} photometry. This is because the long duration of the transit limits us to partial transit baselines when observing with WIRC, which also has a slightly lower photometric precision than \emph{Kepler}.
In \autoref{transit_cornerplot}, we plot the posterior probability distributions for the physically meaningful transit shape parameters from the joint fit (all parameters are global other than ${\frac{R_p}{R_{*}}}$ in the \textit{Kepler} and WIRC bandpasses, and the $t_0$ parameter on each WIRC night). 

\begin{figure}
  \includegraphics[width=8.5cm]{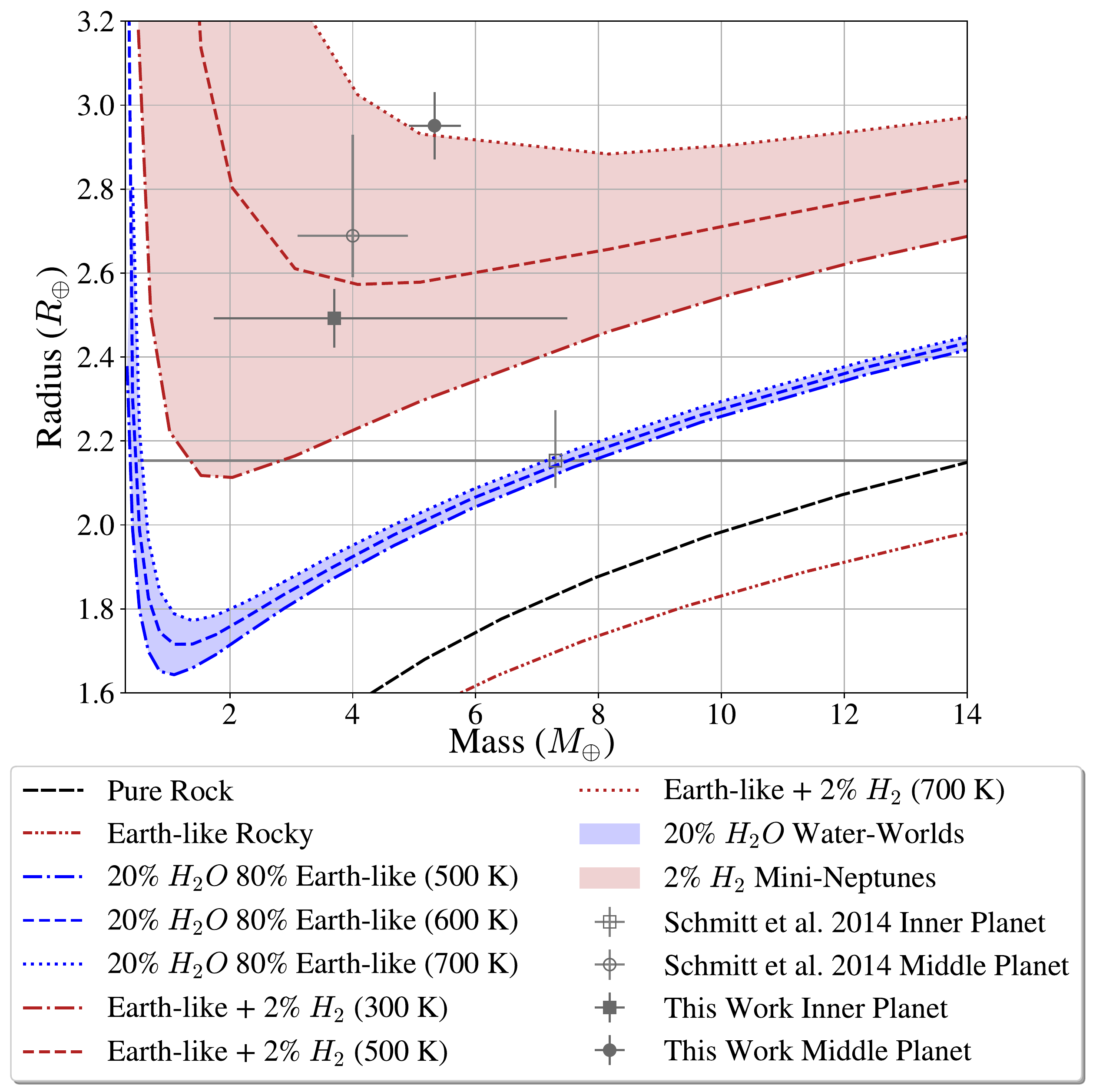}
  \caption{Updated mass and radius measurements for the two inner sub-Neptunes (Kepler-289b and d) compared to the original values from \cite{schmitt}. 
  We overplot constant composition curves for 100\% iron, an earth-like rock-iron mix, irradiated hydrogen-rich atmospheres on top of an Earth-like rocky core, and irradiated water worlds from \cite{zeng_2016}
  }\label{M_R_innerplanets}
\end{figure}

\subsection{TTV Analysis}
We fit our updated set of transit mid-time measurements using the \texttt{TTVfast} package \citep{ttvfast}. \texttt{TTVfast} is a computationally efficient $n$-body code that computes transit times as a function of the planetary masses and orbital elements at a reference epoch, which we chose to be $T_0 = 2455865$ (BJD). We fixed the orbital inclinations to 90$^{\circ}$ and the longitudes of the ascending nodes to 0$^{\circ}$ because our transit profile fits indicate that the inclination dispersion is small, and the TTV solution is second order in mutual inclination \citep{nesvorny_2014, hadden_lithwick_2016}.  We used \texttt{emcee} to map the posterior probability distributions for the masses and orbital elements for this three-planet system. We ran 
the sampler with 50 walkers for $3.5\times\mathrm{10}^{6}$ iterations and discard the initial $5\times\mathrm{10}^{5}$ steps as burn-in. We checked for convergence by verifying that the number of iterations for each parameter in our MCMC chain exceeded 50 autocorrelation lengths as calculated by \texttt{emcee}. Our observed TTVs, as well as 100 random draws of the TTV posterior distribution from our TTVfast fit, are shown in \autoref{ttvplot}, while the corner plot showing the posterior probability distributions for the planet masses and eccentricity vectors is shown in \autoref{ttv_cornerplot}. We use our improved dynamical solution to predict the transit mid-times for all three planets until 2032, which are listed along with our observed transit times in \autoref{times_table}.

\begin{table*}[!htbp]
\caption{All observed transit epochs, mid-times, and uncertainties for the three Kepler-289 planets, as well as the predicted mid-times and associated 1$\sigma$ uncertainties from our best-fit TTV model, extending until January 1$^{st}$, 2032. The table is abbreviated here, but available in full online.}
\centering
\begin{tabular}{p{2cm}p{1.5cm}p{2.3cm}p{1.8cm}p{2.3cm}p{1.8cm}}
\textbf{Planet} & \textbf{Transit Number} & \textbf{Observed Mid-time (JD-2454000)} & \textbf{Error (days)} & \textbf{Predicted Mid-time (JD-2454000)} & \textbf{Error (days)} \\ \hline
Kepler-289b & 0 & 965.7135 & 0.0110 & 965.6898 & 0.0031 \\
Kepler-289b & 1 & - & - & 1000.2328 & 0.0036 \\
Kepler-289b & 2 & - & - & 1034.7792 & 0.0041 \\
Kepler-289b & 3 & 1069.3174 & 0.0036 & 1069.3195 & 0.0054 \\
Kepler-289b & 4 & - & - & 1103.8616 & 0.0056 \\
\end{tabular}
\label{times_table}
\end{table*}

\begin{figure}
  \includegraphics[width=9cm]{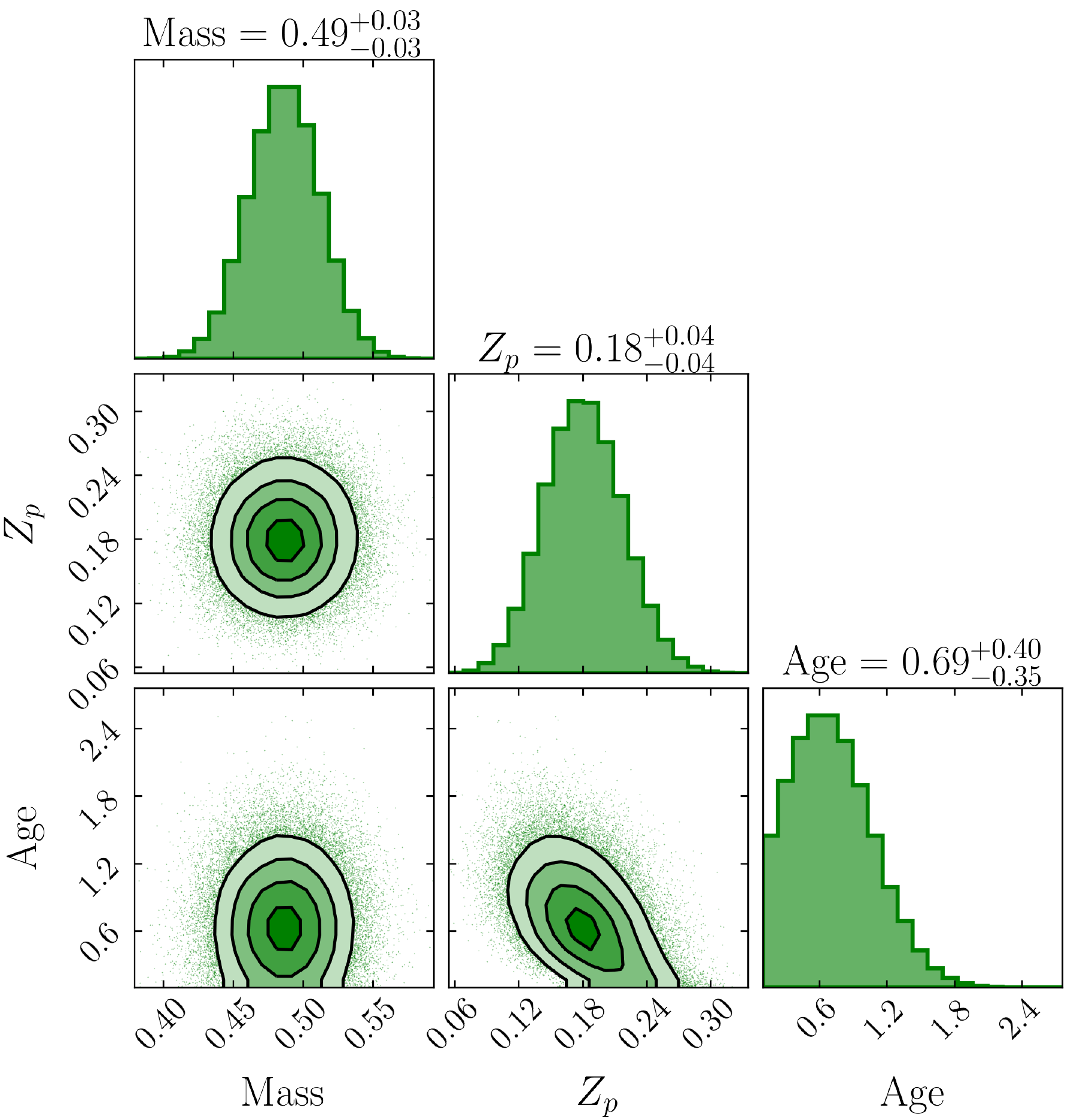}
  \caption{Posterior probability distributions for the bulk metallicity, planetary mass, and age of the outermost planet (Kepler-289c). We use a planetary evolution model as described in \cite{thorngren_2016} to compute the planet's bulk metal content as a function of its observed radius, mass, and age, which are taken from published values in the literature.}
  \label{metallicity}
\end{figure}

\section{Results}\label{sec:results}
Our new analysis decreases the mass uncertainties for all three planets by more than a factor of 2, with the innermost planet seeing the greatest improvement relative to the original constraints found in \cite{schmitt}. We summarize the final planetary masses, radii, densities, and orbital properties from our photometric and TTV analyses in \autoref{system_table}. In this section, we use these values to constrain the bulk densities and possible compositions of the three planets.

We find that Kepler-289b is a 4.1 $M_{\oplus}$ low-density sub-Neptune, as illustrated in \autoref{M_R_innerplanets}. Both the inner and middle planets are significantly larger than a pure-water planet of the same mass. This implies that both planets likely possess a primordial hydrogen- and helium-rich atmosphere, which increases their apparent radii. We use the models in \cite{lopez_fortney_2014} to calculate the core-to-atmosphere mass fraction for a scenario in which we assume that the planets consist of Earth-like rocky cores surrounded by a solar metallicity gas envelope.  We find that the measured densities of the inner and middle planets can be matched by hydrogen-rich envelopes that constitute 3\% and 0.9\% of their total planet masses, respectively. Both of these values are consistent with the estimated atmospheric mass fractions of the broader sub-Neptune population \citep[e.g.,][]{owen_wu_2013, lopez_fortney_2014, eve}. 

\begin{table*}[]
\caption{Orbital and planetary parameters for Kepler-289b, d, and c. Each of these parameters is retrieved from our transit and TTV fits, other than the derived parameters of $M_p$, $R_p$, and $\rho$, which incorporate errors on the stellar mass and radius of $M_{\star}$ = 1.08 $\pm$ 0.02 $M_{\odot}$ and $R_{\star}$ = 1.00 $\pm$ 0.02 $R_{\odot}$ from \cite{schmitt}.}

\centering
\setlength\extrarowheight{-3pt}
\hspace*{-1.3cm}\begin{tabular}{K{2.5cm} K{2.5cm} K{2.5cm} K{2.5cm}}
\multicolumn{4}{c}{\textbf{Kepler-289b}} \\
 \noalign{\vskip .1cm} 
 \hline\hline
 \noalign{\vskip .1cm} 
 & Schmitt et al. 2014 & Kepler Reanalysis & Kepler Reanalysis + Palomar \\
\hline
\noalign{\vskip .1cm} 
$P$ (days) & 34.5450±0.0005 & $34.5383\substack{+0.0006  \\ -0.0006}$ & $34.5383\substack{+0.0006 \\ -0.0006}$ \\
$T_0$ (JD-2454000) & 965.6404$\pm$0.0040 & $965.6880\substack{+0.0030 \\ -0.0032}$ & $965.6879\substack{+0.0006 \\ -0.0006}$\\
$M_p$ $(M_{\oplus})$ & 7.3$\pm$6.8 & $4.39\substack{+4.32 \\ -2.30}$ & $3.70\substack{+3.79  \\ -1.96}$ \\
$R_p$ $(R_{\oplus})$ & 2.15$\pm$0.10 & 2.49$\pm$0.07 & 2.49$\pm$0.07 \\
$\rho$ (g~cm$^{-3}$) & 4.1$\pm$3.9 & $1.56\substack{+1.54  \\ -0.82}$ & $1.45\substack{+1.5  \\ -0.77}$ \\
$R_p$/$R_*$ & $0.0197\substack{+0.0011 \\ -0.0006}$ & $0.0228\substack{+0.0004 \\ -0.0004}$  & $0.0228\substack{+0.0004 \\ -0.0004}$ \\
$M_p$/$M_*$(×10$^{-5}$) & 2.0$\pm$1.9 & $1.22\substack{+1.2 \\ -0.64}$ & $1.13\substack{+1.17 \\ -0.6}$ \\
$a/R_*$ & $71.1\substack{+10 \\ -20}$ & $46.38\substack{+2.1 \\ -2.0}$ & $46.38\substack{+2.1 \\ -2.0}$  \\
$i (^{\circ}$) & $89.59\substack{+0.30 \\ -0.48}$ & $88.98\substack{+0.06 \\ -0.07}$ & $88.98\substack{+0.06 \\ -0.07}$ \\
$e\cos(\omega$) & -0.0215$\pm${0.0255} & - & - \\
$e\sin(\omega$) & -0.0113$\pm${0.0239} & - & - \\
$\sqrt{e}$cos($\omega$) & - & $-0.0364\substack{+0.0802 \\ -0.0756}$ & $-0.0333\substack{+0.0757 \\ -0.0873}$ \\
$\sqrt{e}$sin($\omega$) & - & $-0.1776\substack{+0.0816 \\ -0.0564}$ & $-0.1858\substack{+0.0820 \\ -0.0564}$ \\
\end{tabular}

\centering
\setlength\extrarowheight{-3pt}
\hspace*{-1.3cm}\begin{tabular}{K{2.5cm} K{2.3cm} K{2.3cm} K{2.3cm}}
 \noalign{\vskip .2cm} 
\multicolumn{4}{c}{\textbf{Kepler-289d}} \\
 \noalign{\vskip .1cm} 
 \hline\hline
 \noalign{\vskip .1cm} 
 & Schmitt et al. 2014 & Kepler Reanalysis & Kepler Reanalysis + Palomar \\
\hline
\noalign{\vskip .1cm} 
$P$ (days) & 66.0634$\pm$0.0114 & $66.0281\substack{+0.0048  \\ -0.0042}$  & $66.0282\substack{+0.0044  \\ -0.0039}$ \\
$T_0$ (JD-2454000) & 975.6436$\pm$0.0068 & $975.6436\substack{+0.0053  \\ -0.0053}$ & $975.6240\substack{+0.0051  \\ -0.0052}$ \\
$M_p$ ($M_{\oplus}$) & 4.00$\pm$0.90 & $5.97\substack{+0.50  \\ -0.50}$ & $5.33\substack{+0.43  \\ -0.42}$ \\
$R_p$ $(R_{\oplus})$ & 2.68$\pm$0.17 & 3.03$\pm$0.08 & 3.03$\pm$0.08 \\
$\rho$  (g~cm$^{-3}$) & 1.2$\pm$0.3 & $1.18\substack{+0.1  \\ -0.1}$ & $1.14\substack{+0.09  \\ -0.09}$ \\
$R_p$/$R_\star$ & $0.0246\substack{+0.0022 \\ -0.0009}$ & $0.0270\substack{+0.0005 \\ -0.0005}$ & $0.0270\substack{+0.0005 \\ -0.0005}$  \\
$M_p$/$M_\star$(×10$^{-5}$) & 1.1$\pm$0.2 & $1.66\substack{+0.14 \\ -0.14}$ & $1.6\substack{+0.13 \\ -0.13}$ \\
$a/R_\star$ & $117.8\substack{+21 \\ -42}$ & $70.52\substack{+2.95 \\ -2.85}$ & $70.52\substack{+2.95 \\ -2.85}$ \\
$i (^{\circ}$) & $89.73\substack{+0.20 \\ -0.38}$ & $89.31\substack{+0.04 \\ -0.04}$ & $89.31\substack{+0.04 \\ -0.04}$ \\
$e\cos(\omega$) & -0.0035$\pm${0.0022} & - & - \\
$e\sin(\omega$) & -0.0108$\pm${0.0122} & - & - \\
$\sqrt{e}\cos(\omega$) & - & $0.0340\substack{+0.0164  \\ -0.0208}$ & $0.0293\substack{+0.0174  \\ -0.0225}$ \\
$\sqrt{e}\sin(\omega$) & - & $-0.0164\substack{+0.0537  \\ -0.0498}$ & $-0.0096\substack{+0.0501  \\ -0.0500}$ \\
\end{tabular}

\centering
\setlength\extrarowheight{-3pt}
\hspace*{-1.3cm}\begin{tabular}{K{2.5cm} K{2.3cm} K{2.3cm} K{2.3cm}}
 \noalign{\vskip .2cm} 
\multicolumn{4}{c}{\textbf{Kepler-289c}} \\
 \noalign{\vskip .1cm} 
 \hline\hline
 \noalign{\vskip .1cm} 
 & Schmitt et al. 2014 & Kepler Reanalysis & Kepler Reanalysis + Palomar \\
\hline
\noalign{\vskip .1cm} 
$P$ (days) & 125.8518$\pm$0.0076 & $125.8727\substack{+0.0036  \\ -0.0022}$ & $125.8723\substack{+0.0035  \\ -0.0021}$ \\
$T_0$ (JD-2454000) & 1069.6528$\pm$0.0077 & $1069.6744\substack{+0.0035  \\ -0.0023}$ & $1069.6734\substack{+0.0035  \\ -0.0021}$ \\
$M_p (M_{\oplus})$ & 132$\pm$17 & $170.35\substack{+7.95  \\ -8.67}$ & $157.18\substack{+7.04  \\ -7.31}$\footnote{In units of $M_{Jup}$, Kepler-289c's mass is 0.49 $\pm$ 0.02} \\
$R_p (R_{\oplus})$ & 11.59$\pm$0.19 & 11.31$\pm$0.23 & 11.23$\pm$0.21 \\
$\rho$ (g~cm$^{-3}$) & 0.47$\pm$0.06 & $0.65\substack{+0.03  \\ -0.03}$ & $0.66\substack{+0.03  \\ -0.03}$ \\
$R_p$/$R_\star$ & $0.10620\substack{+0.00049 \\ -0.00050}$ & $0.10373\substack{+0.00040 \\ -0.00038}$ &  $0.10297\substack{+0.00027 \\ -0.00026}$  \\
$M_p$/$M_\star$(×10$^{-5}$) & 36.43$\pm$4.66 & $47.36\substack{+2.21 \\ -2.41}$ & $47.19\substack{+2.11 \\ -2.23}$ \\
$a/R_\star$ & 109.5$\pm$1.2 & $104.7\substack{+5.9 \\ -5.1}$ & $108.95\substack{+0.75 \\ -0.75}$ \\
$i$ ($^{\circ}$) & $89.794\substack{+0.017 \\ -0.016}$ & $89.74\substack{+0.07 \\ -0.05}$ & $89.78\substack{+0.04 \\ -0.04}$ \\
$e\cos(\omega$) & 0.0032$\pm${0.0066} & - & - \\
$e\sin(\omega$) & 0.0033$\pm${0.0086} & - & - \\
$\sqrt{e}$cos($\omega$) & - & $0.1160\substack{+0.0152  \\ -0.0187}$ & $0.1114\substack{+0.0164  \\ -0.0194}$ \\
$\sqrt{e}$sin($\omega$) & - & $0.0277\substack{+0.0386  \\ -0.0382}$ & $0.0399\substack{+0.0278  \\ -0.0350}$ \\
\end{tabular}
\label{system_table}
\end{table*}

\begin{figure}
  \includegraphics[width=8.5cm]{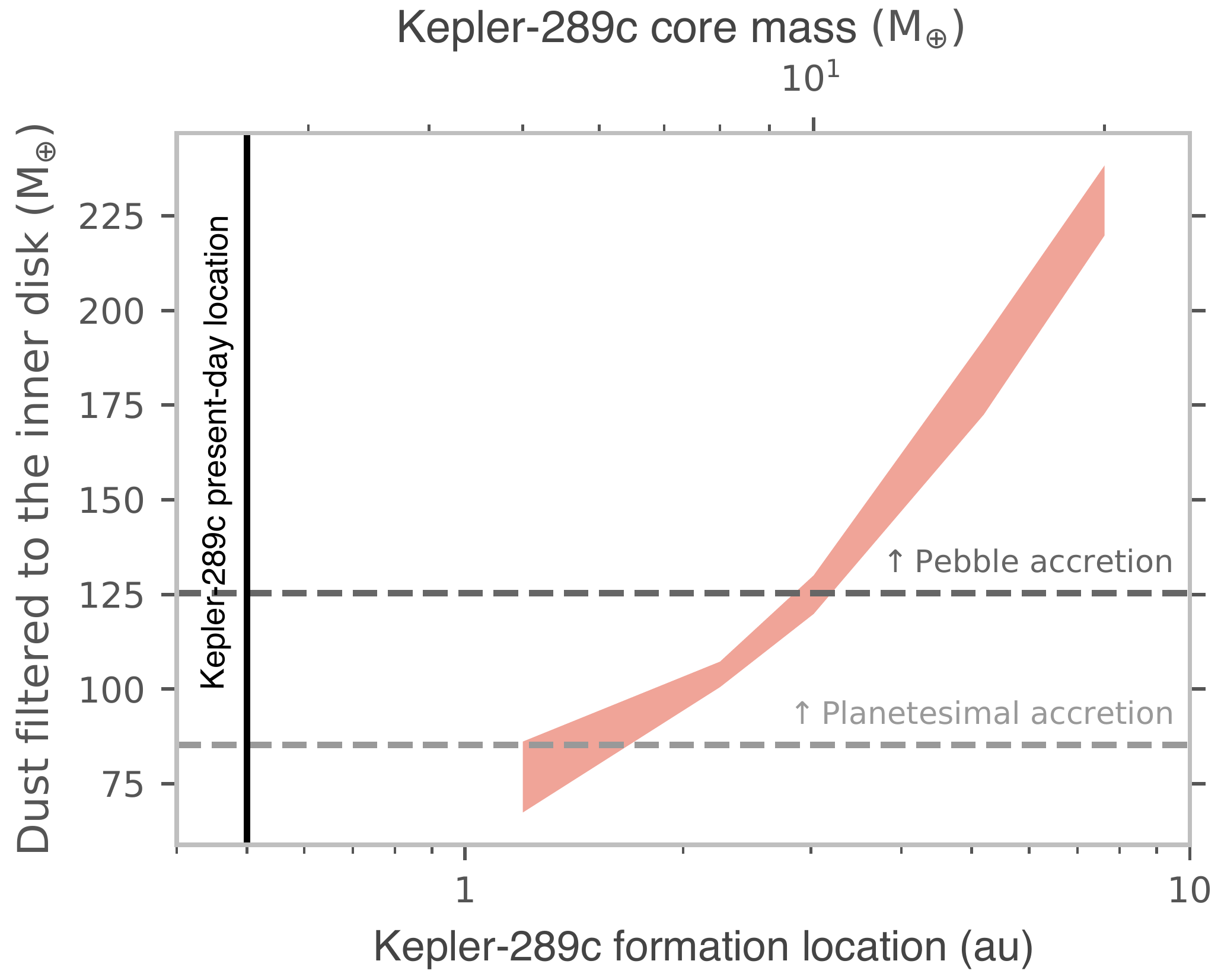}
  \caption{Total dust mass that reaches the inner disk as a function of the assumed giant planet pebble isolation (core) mass (top x axis), and its corresponding formation location (bottom x axis, this is set by the assumed isolation mass).  The dust mass is calculated using the disk models from \cite{Chachan_2022}, where the red shading denotes the spread in predictions caused by varying the assumed disk radii, masses, and dust-to-gas ratios. The minimum estimated dust masses required to form the two inner planets in the planetesimal (grey dashed line) and pebble accretion (black dashed line) are overplotted. Kepler-289c's current orbital semi-major axis is shown as a solid black line.}
  \label{pebble_isolation_plot}
\end{figure}

For the outermost planet, Kepler-289c, we find an updated mass of 0.49 $\pm{0.02}$ $M_{Jup}$. We use a giant planet interior structure and evolution model \citep{thorngren_2016, thorngren_and_fortney_2019} to calculate the corresponding bulk metallicity as a function of the planet's mass, age, and radius. We allow these three parameters to vary freely in the fit with priors based on their measured values from this study and \cite{schmitt}, and show the resulting posterior probability distributions in \autoref{metallicity}. 
For more details on the giant planet structure model, see \cite{thorngren_and_fortney_2019}. We find that this planet has a bulk metallicity of 0.18 $\pm$ 0.04, significantly higher than its approximately solar metallicity host star.  This corresponds to a bulk metal content of 30.5 $\pm$ 6.9 $M_{\oplus}$.  We compare Kepler-289c to the broader population of giant planets characterized in \cite{thorngren_2016} and find that its bulk metallicity is slightly lower than average based on the best-fit mass-metallicity trend model, but still consistent with the overall population distribution. 

Core accretion models for gas giant formation indicate that it is relatively difficult to form planets like Kepler-289c in-situ on close-in orbits \citep{dawson_2018, Poon_2021}. It is therefore possible that Kepler-289c formed farther out and then migrated inward.  Observational studies suggest that the gas giant planet population peaks at separations between $1-10$ au \citep{Fulton_2021}, much farther out that Kepler-289c's present-day orbital separation of $\sim$0.5 au. Although the sample of outer gas giants with inner super-Earths is much smaller, observational studies find that the gas giants in these systems are also typically located on relatively wide orbits \citep{zhu_wu_2018,bryan_2019}. Kepler 289c's lower-than-average bulk metallicity also suggests a more distant formation location, as dust evolution models indicate that the dust opacity should be lower (and the corresponding gas accretion rates for a fixed core mass should be higher) beyond the ice line \citep{Chachan_2021}. 

The fact that both planet pairs in this system have a normalized distance to resonance $\Delta < 0.05$ is also consistent with the hypothesis that they might have migrated inward through the 2:1 resonance at some point in the past \citep[e.g.,][]{weiss_2022}, although the current observational constraints strongly disfavor a resonant configuration. \cite{schmitt} calculated the 3-body Laplace resonance width for this system, and found that both pairs of planets are many tens of widths away from resonance. We empirically estimate the three-body Laplace resonance angle using the method from \cite{mills_2016} and find that it circulates from 0$^{\circ}$–360$^{\circ}$ over the course of the \emph{Kepler} observations, with no indication of libration. We therefore conclude that the Kepler-289 planets are not currently in an orbital resonance, in agreement with \cite{schmitt}. This might be explained by post-nebular dynamical instabilities, which can nudge resonant systems into configurations that are either interior to or exterior to the original resonance \citep{Goldberg_2022}.

\subsection{Constraining the formation location of Kepler-289c}
We next explore whether or not we can place more quantitative constraints on the likely formation location of Kepler-289c using a pebble accretion model as described in \cite{Chachan_2022}. This model assumes that the material available in the inner disk alone is not sufficient to form a system of super-Earths, but must instead be enhanced by a flow of small solids migrating in from the outer disk.
When the outer giant planet core reaches the pebble isolation mass, it cuts off this flow of solids. \cite{Chachan_2022} used disk evolution models to show that the integrated dust mass that reaches the inner disk in this scenario is primarily a function of the giant planet's formation location and corresponding isolation mass, with more distant formation locations (higher isolation masses) allowing more material to reach the inner disk. 

As discussed in the previous section, the masses of the two inner planets in the Kepler-289 system are dominated by solids. This means that they must have accreted a combined total of approximately $\sim$10 $M_{\oplus}$ of solids. If we compare this to the solid mass budget for the inner disk, using the minimum mass extrasolar nebula \citep{Chiang_2013} to estimate the solid content within the orbit of Kepler-289c yields a total solid mass of $\sim$8 $M_{\oplus}$. Clearly, some material must have been transported to the inner disk to form Kepler-289b and d. If this material predominately came from beyond the ice line, approximately 50\% of the original solid mass would have been lost to evaporation when the pebbles crossed the water ice line. The remaining solids must accrete into larger bodies to avoid migrating all the way onto the star. \cite{Drazkowska_2016} combined dust evolution models with planet formation models to show that dust is converted into planetesimals with an efficiency of $\sim$23\% in the inner disk. This implies that 85 $M_{\oplus}$ of solids must have been delivered to the inner disk in order to form the two inner planets in the Kepler-289 system.  If the inner super-Earths formed via pebble accretion, the efficiency with which solids are accreted is somewhat lower due to Kepler-289's large stellar mass \citep[see discussion in \S~5.3 and Appendix A of ][]{Chachan_2022}, resulting in a higher mass threshold of approximately 125 $M_{\oplus}$.

Since the total dust mass delivered to the inner disk is a function of the giant planet's formation location, we can use the corresponding pebble isolation mass for a given formation location to place a lower limit on the initial orbital distance of Kepler-289c. We run a new grid of disk models that account for Kepler-289's mass and its expected protostellar luminosity but is otherwise identical to models presented in \cite{Chachan_2022}. This grid spans a range of assumed disk radii, solid masses, and dust-to-gas ratios, and we only consider the subset of disks where the giant planet core reaches the isolation mass in less than 1 Myr.  We show the resulting solid fluxes as a function of the giant planet's formation location and corresponding pebble isolation mass in \autoref{pebble_isolation_plot}.  In these models, a giant planet core located just outside 1 au, which has a corresponding pebble isolation mass of $\sim$5 $M_{\oplus}$, allows $\sim$85 $M_{\oplus}$ to be delivered to the inner disk.  This is enough material to form the inner super-Earths via planetesimal accretion, but such a small core is unlikely to accrete enough gas to become a gas giant \citep[e.g.,][]{Rafikov_2006, Lee_2019}.  A core located just outside 3 au would have an isolation mass of 10 $M_{\oplus}$, making it easier to accrete a massive gas envelope and also providing enough solids to the inner disk to exceed the higher pebble accretion threshold of 125 $M_{\oplus}$.  A formation location beyond 3 au would also be consistent with the semi-major axis distribution of the broader population of giant planets measured in \cite{Fulton_2021}, which peaks at 3.6$^{+2.0}_{-1.8}$~au.  

Kepler-289c has a total metal content of $\sim$30 $M_{\oplus}$ from our bulk metallicity constraints, but if it has a structure similar to that of Jupiter and Saturn it is likely that a significant fraction of these solids are distributed throughout its interior \citep{Wahl_2017, Mankovich_2021}.  Such a distribution might plausibly be attributed to the continued accretion of large solids (planetesimals) after the giant planet core reaches the isolation mass \citep[see][for a review]{Helled_2022}, allowing for a core mass that is less than the planet's total estimated solid mass. If a larger fraction of Kepler-289c's solids originated from its core, it would imply a more distant formation location and a correspondingly larger flux of solids to the inner disk.  We therefore conclude that 3 au is a reasonable lower limit for the likely formation location of Kepler-289c.

\section{Conclusions}\label{sec:conclusions}

In this study, we used transit timing variations to obtain improved mass measurements for the three planets in the Kepler-289 system, which have orbital period ratios of 1.9:1 for each adjacent pair of planets. We re-analyzed the \emph{Kepler} photometry for this system and observed two new transits of the outermost planet, Kepler-289c, with WIRC on the 200" Hale Telescope at Palomar Observatory. Our re-analysis of the \emph{Kepler} photometry allowed us to better correct for the star's high photometric variability, resulting in modestly improved radius constraints and transit times.  Our Palomar transit observations extended the original four-year \emph{Kepler} TTV baseline by an additional 7.5 years, allowing for an improved dynamical solution.  Our updated mass uncertainties for all three planets are more than a factor of two smaller than the original values reported by \cite{schmitt}.

We use our improved mass and radii constraints to establish that the sub-Neptune sized inner and middle planets, Kepler-289b and d, are both low-density (1.45 and 1.14 g~cm$^{-3}$) planets. If we assume that these planets both have Earth-like cores with solar metallicity envelopes, we calculate that they should have atmospheric mass fractions of 3.0\% and 0.9\%, respectively.  This is consistent with their location on the upper side of the radius valley. Since Kepler-289 is relatively young, it is possible that these two planets may host present-day atmospheric outflows.  Although this system is too faint to measure helium outflow signatures with current ground-based telescopes, it might be accessible to upcoming thirty-meter class telescopes.

We calculate the three-body Laplace resonance angles for the system and find that Kepler-289 is not in the 1:2:4 Laplace orbital resonance, in agreement with \cite{schmitt}. We argue that this system might have migrated into a resonant chain, which was later disrupted by dynamical instabilities. This is consistent with both observations and models of giant planet formation, which suggest that planets like Kepler-289c (0.5 $M_{Jup}$, 0.5 au present-day location) form more readily outside 1 au.  We combine our updated planet masses with disk evolution models to constrain the formation location of Kepler-289c, and conclude that it likely originated at or beyond 3~au. 

Transiting, dynamically interacting multi-planet systems provide us with a wealth of information about planet formation and migration processes.  Kepler-289's unique architecture makes it a particularly valuable test case for exploring the connections between close-in sup-Neptune planets and outer gas giant companions.  Although its relative faintness also makes it a challenging target for the \emph{James Webb Space Telescope}, atmospheric composition studies of Kepler-289c with next-generation space telescopes could provide additional constraints on its formation location \citep{Madhusudhan_2019}.

\section{Acknowledgements}
We thank the Palomar Observatory telescope operators, support astronomers, and directorate for their support of this work, especially Kajse Peffer, Kevin Rykoski, Carolyn Heffner, Andy Boden, and Tom Barlow. Part of this program is supported by JPL Hale telescope time allocation. Some of the data presented in this paper were obtained from the Mikulski Archive for Space Telescopes (MAST). STScI is operated by the Association of Universities for Research in Astronomy, Inc., under NASA contract NAS5-26555. Support for MAST for non-HST data is provided by the NASA Office of Space Science via grant NNX13AC07G and by other grants and contracts. H.A.K. acknowledges support from NSF CAREER grant 1555095. SV is supported by an NSF Graduate Research Fellowship. This research made use of Lightkurve, a Python package for Kepler and TESS data analysis \citep{lightkurve}. This research made use of \texttt{exoplanet} \citep{exoplanet:joss} and its dependencies \citep{exoplanet:agol20,
exoplanet:arviz, exoplanet:astropy13, exoplanet:astropy18, exoplanet:kipping13,
exoplanet:luger18, exoplanet:pymc3, exoplanet:theano}.

\facilities{\textit{Kepler}, Palomar: Hale 200`` Telescope}

\software{astropy \citep{astropy},
          scipy \citep{scipy},
          numpy \citep{scipy},
          matplotlib \citep{matplotlib}
          rebound \citep{rebound}, 
          BATMAN \citep{batman},
          emcee \citep{emcee},
          corner \citep{corner},
          \texttt{exoplanet} \citep{exoplanet:joss}
          }
          
\bibliography{references}{}
\bibliographystyle{aasjournal}

\begin{appendix}

\section{Transit and TTV Model Plots and Posterior Distributions}

\begin{figure}[!hb]
  \includegraphics[width=17.5cm]{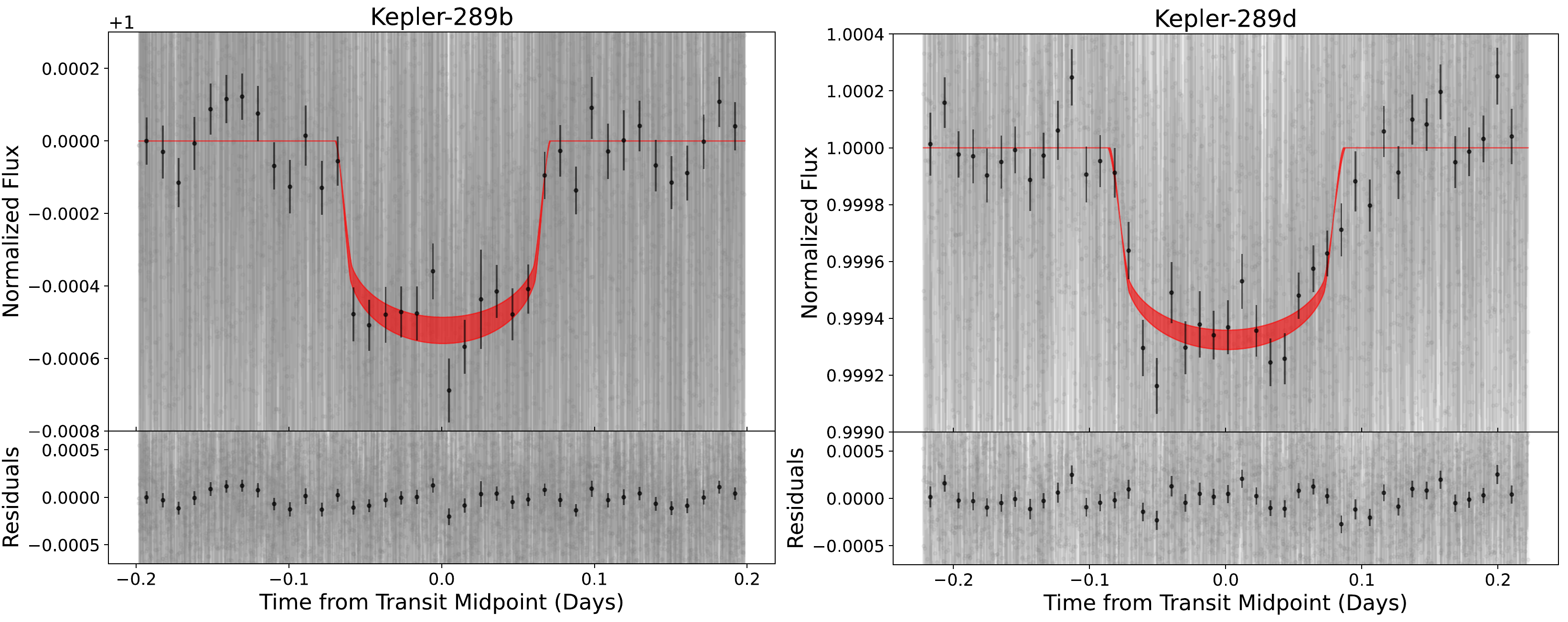}
  \caption{Gray points show the phased \emph{Kepler} data. The left and right panels are the inner and middle planets, respectively. Black points are binned to 15 minutes. The red shading indicates the middle 68\% range from our posterior distribution.}
  \label{Kepler289b_transit_plot}
\end{figure}

\begin{figure}[!hb]
  \includegraphics[width=\textwidth]{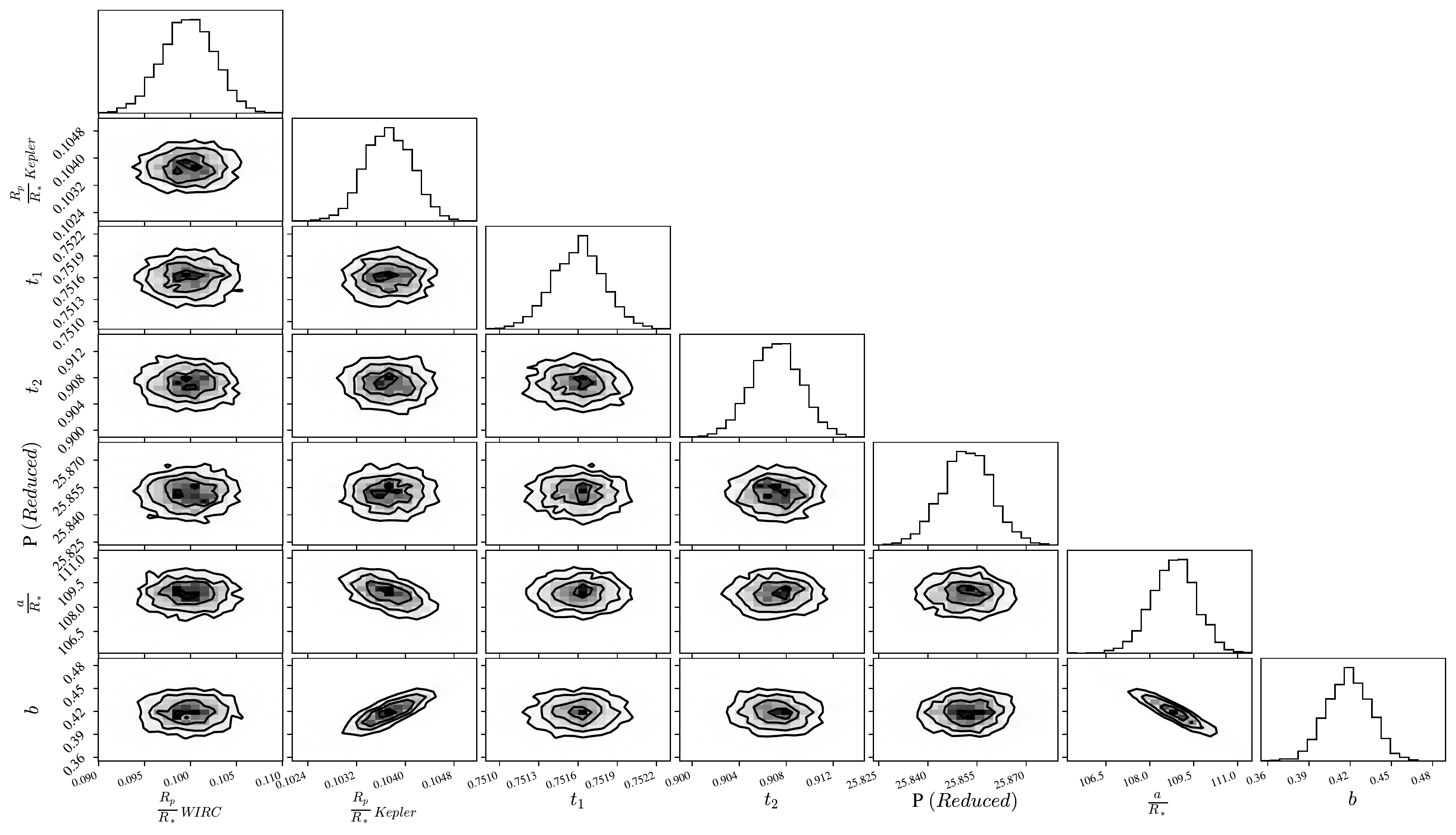}
  \caption{Posterior probability distributions for the transit shape parameters from the MCMC joint fit to the \emph{Kepler} and WIRC data. For ease of viewing in the plot, we label the period minus 100 days as the reduced period. $t_1$ is the BJD transit center time of the first WIRC night - 2458719 days, and $t_2$ is the BJD transit center time of the second WIRC night - 2459474 days.}
  \label{transit_cornerplot}
\end{figure}

\begin{figure*}[!htb]
  \includegraphics[width=\textwidth]{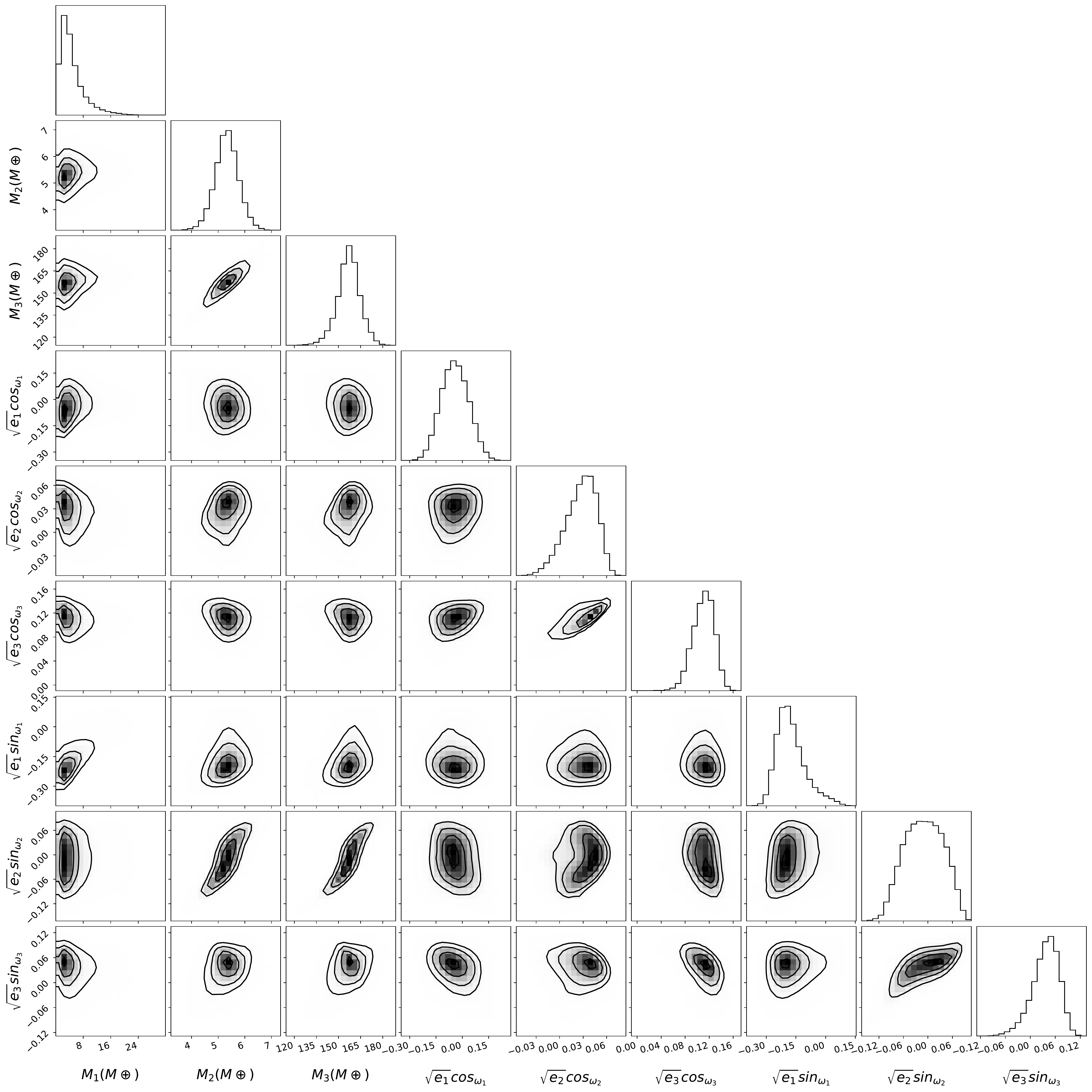}
  \caption{Posterior probability distributions for the planet masses and eccentricity vectors from the TTV fit. We also allowed the orbital periods and initial transit times for each planet to vary as free parameters in this fit, but we omit them from this plot for ease of viewing.}
  \label{ttv_cornerplot}
\end{figure*}

\clearpage

\end{appendix}

\end{document}